\documentclass{mem}
\usepackage{natbib}\usepackage{txfonts}\usepackage{balance}
\usepackage{graphicx}
\usepackage[a4paper,breaklinks,dvipdfm]{hyperref}
\idline{75}{282}
\begin{document}
\def\teff{$T\rm_{eff }$}
\def\kms{$\mathrm {km s}^{-1}$}

\title{Cosmological distance indicators by
coalescing binaries
}


\author{
M. De Laurentis\inst{1,2}, S. Capozziello\inst{1,2}, I. De Martino\inst{3}, M. Formisano\inst{4}}

  \offprints{M. De Laurentis}

\institute{
Dipartimento di Scienze Fisiche, Universit\`a di Napoli "Federico II"
\and
INFN sez. di Napoli Compl. Univ. di Monte S. Angelo, Edificio G, Via Cinthia, I-80126 - Napoli, Italy
\and
Departamento de  Fisica Teorica, Universidad de Salamanca, 37008 Salamanca, Spain
\and
Dipartimento di Fisica, Universit\`a di Roma "La Sapienza", Piazzale Aldo
Moro 5, I-00185 Roma, Italy
}

\authorrunning{M. De Laurentis }

\titlerunning{Cosmological distance indicators by
coalescing binaries}

\abstract{
Gravitational waves detected  from well-localized inspiraling
binaries would allow  to  determine, directly and independently,
both   binary  luminosity  and  redshift. In this case, such
systems could behave as "standard candles"  providing  an
excellent probe of cosmic distances up to $z <0.1$ and thus
complementing other indicators of cosmological distance ladder.

}

\maketitle{}

\section{Introduction}
Coalescing binaries systems are usually considered strong
emitter of gravitational waves (GW), ripples of space-time due to
the presence of accelerated masses  in analogy with the
electromagnetic waves, due to accelerated charged. The coalescence
of astrophysical systems containing  relativistic  objects as
neutron stars (NS), white dwarves (WD) and black holes (BH)
constitute very standard GW sources which could be extremely
useful for cosmological distance ladder if physical features of GW
emission are well determined.
These binaries systems, have  components
that are gradually inspiralling one over the other   as the result
of energy and angular momentum loss due to (also) gravitational
radiation. As  a consequence the GW frequency is increasing and,
if observed, could constitute a "signature" for the whole system
dynamics. The coalescence of a compact binary system is usually
classified in three stages, which are not very well delimited one
from another, namely the \emph{inspiral phase}, the \emph{merger
phase} and the \emph{ring-down phase}.
Temporal
interval between the inspiral phase and the merger one is called
\emph{coalescing time}, interesting for detectors as the American
LIGO (Laser Interferometer Gravitational-Wave Observatory)
\citep{LIGO} and French/Italian VIRGO \citep{VIRGO}.
A remarkable fact about
binary coalescence is that it can provide an {\it absolute
measurement} of the source distance: this is an extremely
important event in Astronomy \citep{original}.
In fact, here we want to show that such systems could be used as reliable
standard candles.

\section{Coalescing binaries as standard candles}

The fact that the binary coalescence can provide an absolute measurement of the distance to the source, can be understood looking at the waveform of an inspiraling binary; as long as the system is not at cosmological distances (so that we can neglect the expansion of the Universe during the propagation of the wave from the source to the observer) the waveform of the GW, to lowest order in $v/c$ is (see \citep{Maggiore} for a detailed exposition of GW theory)
\begin{equation}
\begin{array}{l}
 h_ +  \left( t \right) =\mathcal{A} \frac{1}{r}\left( {\frac{{\pi f\left( {t_{R} } \right)}}{c}} \right)^{2/3} \left( {\frac{{1 + \cos ^2 i}}{2}} \right)\cos \left[ {\Phi \left( {t_R } \right)} \right]\,
 \end{array}
\end{equation}

\begin{equation}
\begin{array}{l}
h_ \times  \left( t \right) = \mathcal{A}  \frac{1}{r}\left( {\frac{{\pi f\left( {t_{R} } \right)}}{c}} \right)^{2/3} \cos i\sin \left[ {\Phi \left( {t_{R} } \right)} \right]\\
  \end{array}
\label{eq:polariz}
\end{equation}

where $\mathcal{A}=4\left( \frac{GM_C }{c^2} \right)^{5/3} $,
$h_+$ and $h_{\times}$ are the amplitudes for the two polarizations of the GW, 
and  $i$ is the inclination of the orbit with respect to the line of
sight, 
\begin{equation}
M_c =\frac{(m_1m_2)^{3/5}}{(m_1+m_2)^{1/5}}
\end{equation}
is a combination of the masses of the two stars known as the chirp mass,
and 
$r$ is the distance to the source; $f $ is the frequency of the GW, which
evolves in time according to 
\begin{equation}
\dot f = \frac{{96}}{5}\pi ^{8/3} \left( \frac{GM_C}{c^3 } \right)^{5/3} f^{11/3}\, ,\\
\label{eq:frequenza}
\end{equation}
$t$ is retarded time, and the phase $\Phi$ is given by 
\begin{equation}
\Phi (t) = 2\pi \int^t_{t_0} dt' \, f (t')\, .
\end{equation}
For a binary at a  cosmological distance, i.e. 
at redshift $z$,  taking into account the propagation in a
Friedmann-Robertson-Walker Universe, 
these equations are modified in a very simple way:
\begin{enumerate}
\item The frequency that appears in the above formulae is
the frequency measured by the observer, $f_{\rm obs}$, which is red-shifted
with respect to the source frequency $f_s$,
i.e. $f_{\rm obs}=f_s/(1+z)$, and
similarly $t$ and $t_{ret}$ are measured with the observer's clock.
\item The chirp mass $M_c$ must be replaced by
${\cal M}_c =(1+z) M_c$.
\item The distance $r$ to the source must be replaced by the luminosity distance 
$d_L(z)$.
\end{enumerate}
Then, the signal received by the observed from
a binary inspiral at redshift $z$, when expressed in terms of the observer
time $t$ , is given by

\begin{equation}
h_ +  \left( t \right) = h_c \left( {t_{R} } \right)\frac{{1 + \cos ^2 i}}{2}\cos \left[ {\Phi \left( {t_{R} } \right)} \right]\, ,
\end{equation}

\begin{equation}
h_ \times  \left( t \right) = h_c \left( {t_{R} } \right)\cos i\sin \left[ {\Phi \left( {t_{R} } \right)} \right]\, .
\end{equation}
where
\begin{equation}
h_c \left( t \right) = \frac{4}{{d_L \left( z \right)}}\left( {\frac{{G {\cal{M_C}} \left( z \right)}}{{c^2 }}} \right)^{5/3} \left( {\frac{{\pi f \left( t \right)}}{c}} \right)^{2/3}\, ,
\label{eq:Dis}
\end{equation}

Let us recall that the luminosity distance $d_L$ of a source
is defined by
\begin{equation}
{\cal F}=\frac{\cal L}{4\pi d_L^2}\, ,
\end{equation}
where ${\cal F}$ is the  flux
(energy per unit time per unit area)
measured by the observer, and  ${\cal L}$ is 
the absolute luminosity of the source,
i.e. the power that it radiates in its rest frame.
For small redshifts, $d_L$ is related to the present value of the
Hubble parameter $H_0$ and to the deceleration parameter $q_0$  by
\begin{equation}
\frac{H_0d_L}{c}=z+\frac{1}{2}(1-q_0) z^2+\ldots\, .
\end{equation}
The first term of this expansion give just the Hubble law
$z\simeq (H_0/c) d_L$, which states that
redshifts are proportional to  distances. The term $O(z^2)$ is the
correction to the linear law for moderate redshifts. 
For large redshifts, the
Taylor series is no longer appropriate, and the whole expansion history of
the Universe is encoded in a function $d_L(z)$. As an example, for a 
spatially flat
Universe, one finds
\begin{equation}
d_L(z)=c \, (1+z)\,\int_0^z\, \frac{dz'}{H(z')}\, ,
\end{equation}
where $H(z)$ is the value of the Hubble parameter at redshift $z$. 
Knowing $d_L(z)$ we can therefore obtain $H(z)$.
This shows that the luminosity distance function
$d_L(z)$ is an extremely important
quantity, which encodes the
whole expansion history of the
Universe. 

Now we can understand why coalescing binaries are standard candles.
Suppose that 
we can measure  the amplitudes of both polarizations 
$h_+,h_{\times}$, as well as  $\dot{f}_{\rm obs}$ 
(for  ground-based interferometers, this actually
requires correlations between different detectors). The amplitude of
$h_+$ is $h_c (1+\cos^2\iota)/2$, while the amplitude of
$h_{\times}$ is $h_c\cos\iota$. From their ratio we can therefore obtain the
value of $\cos\iota$, that is, the inclination of the orbit 
with respect to the line of
sight. 
On the other hand, (\ref{eq:frequenza}) (with the replacement $M_c\rightarrow {\cal M}_c$
mentioned abobe) shows that if we measure 
the value of $\dot{f}_{\rm obs}$ corresponding to a given value of 
$f_{obs}$, we get ${\cal M}_c$.
Now in the expression for $h_+$ and $h_{\times}$ 
all parameters  have been fixed, except $d_L(z)$.\footnote{It is important
 that the
  ellipticity of the orbit does not enter; it can in fact be shown that, 
by the time that the stars approach the
coalescence stage, angular momentum losses have circularized the orbit to great
accuracy.}
This means that, from the measured
value of $h_+$ (or of $h_{\times}$) we can now read $d_L$. 
If, at the same time, we can measure the redshift $z$ of the source, we have
found a gravitational standard candle, and we can use it to measure the Hubble
constant and, more generally, the evolution of the Universe \citep{Schutz0}.
The difference between gravitational standard candles and
the "traditional" standard candles is that the luminosity distance
is directly linked to the GW polarization  and  there is no
theoretical uncertainty on its determination a part  the redshift
evaluation. Several possibilities have been proposed. Among these
there is the possibility to see an optical counterpart. In
fact, it can be  shown that observations of the GWs emitted by
inspiralling  binary compact systems can be a powerful probe at
cosmological scales. In particular, short GRBs appear related to
such systems and  quite promising as potential GW standard sirens
\citep{noi}\citep{Dalal}). On the other hand, the redshift of the binary
system can be associated to the barycenter of the host galaxy or
the galaxy cluster as we are going to do here.

\section{Numerical results}

We have simulated several coalescing binary systems at redshifts
$z < 0.1$  In this analysis, we do not consider systematic
errors and errors on redshifts to obviate the absence of a
complete catalogue of such systems. The choice of low redshifts is
due to the observational limits of ground-based interferometers
like VIRGO or LIGO.
Some improvements
 are achieved, if we take into account  the future generation of these
 interferometers as
 Advanced VIRGO\footnote{\cite{advirgo}} and Advanced LIGO\footnote{\cite{adligo}}. Advanced VIRGO  is a major upgrade, with the
 goal of increasing  the sensitivity by about one order of magnitude with
 respect to VIRGO in the whole detection band. Such a detector,
 with Advanced LIGO, is expected to see many events every year
 (from 10s to 100s  events/year). 
 In the simulation presented here,
 sources are slightly out of LIGO-VIRGO band but observable, in principle, with  future interferometers.
 
Here, we have used the redshifts taken by NED\footnote{NASA/IPAC  EXTRAGALACTIC  DATABASE} \citep{Abell},
and we have fixed the redshift  using  $z$ at the
barycenter of the host galaxy/cluster, and the binary
\emph{chirp mass} $M_C$,  typically measured, from the Newtonian
part of the signal at upward frequency sweep, to $\sim 0.04\%$ for
a NS/NS binary 
 \cite{Cutler,original}. The distance to the binary $d_L$
("luminosity distance" at cosmological distances) can be inferred,
from the observed waveforms, to a precision $\sim 3/\rho \lesssim
30\%$, where $\rho = S/N$ is the amplitude signal-to-noise ratio
in the total LIGO network (which must exceed about 8 in order that
the false alarm rate be less than the threshold for detection). In
this way, we have fixed  the characteristic amplitude of GWs, and
frequencies are tuned in a range compatible  with such a  fixed
amplitude, then the error on distance luminosity is calculated by the error on the chirp mass with standard error propagation.

\begin{figure}[!h]
\includegraphics[scale = 0.23]{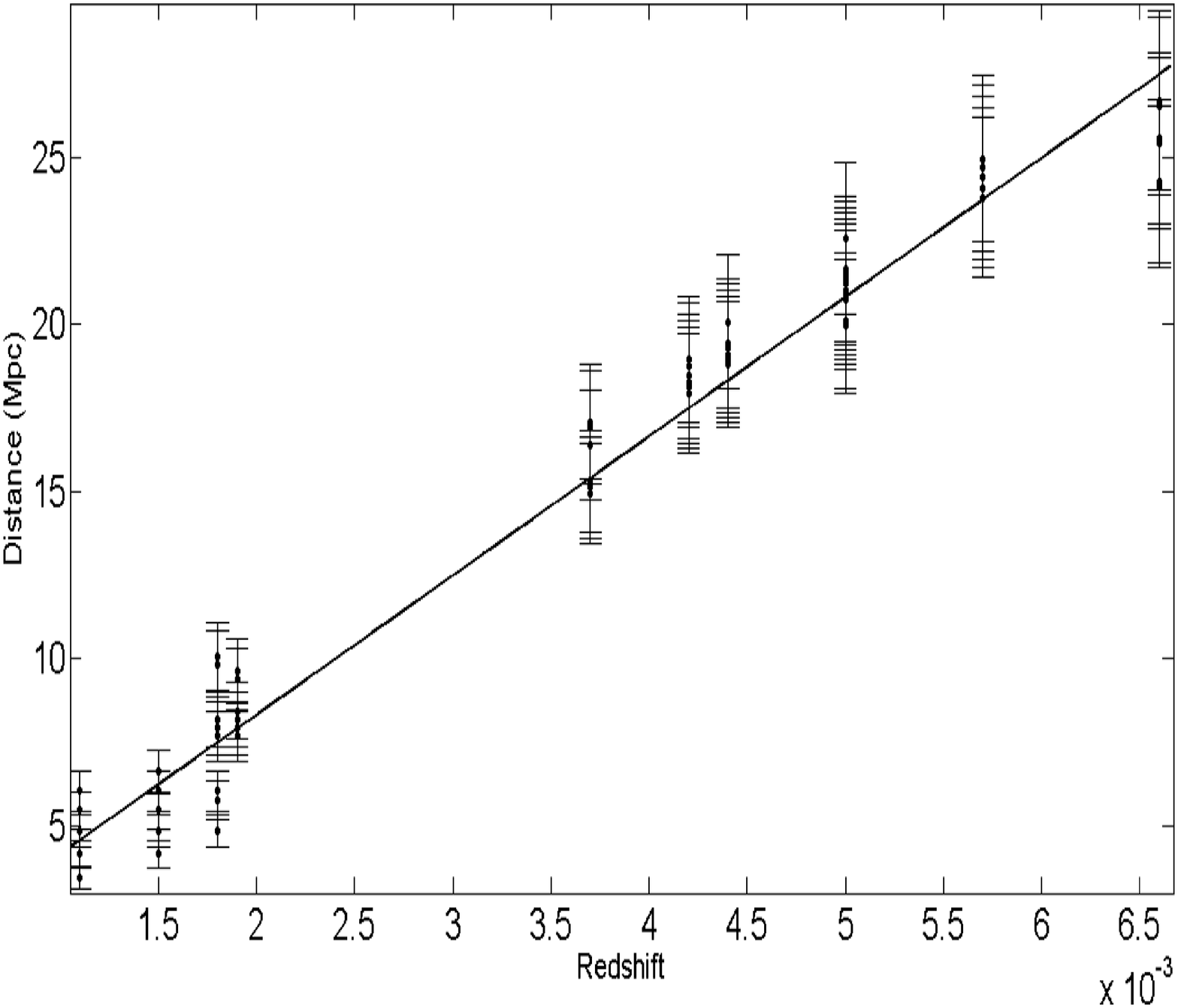}
\caption{Luminosity distance vs redshift for simulated NS-NS systems.}
\label{fig:neutroni}
\end{figure}

The systems considered are NS-NS and BH-BH. For each of them, a particular
frequency range and a characteristic amplitude (beside the chirp
mass) are fixed. We start with the analysis of NS-NS systems ($M_C
= 1.22 M_{\odot}$) with characteristic amplitude fixed to the
value $10^{-22}$. In Table \ref{tab:tabella1}, we report the
redshift, the value of $h_C$ and the frequency range of systems
analyzed.In Fig. \ref{fig:neutroni}, the derived Hubble relation is
reported.

\begin{table}[htbp]
\centering{
\footnotesize{\begin{tabular}{|c|c|c|c|}
  \hline
  \textbf{Object} & \textbf{z} &
  \textbf{$h_c$} & \textbf{Freq.} \\
  & & &  (Hz)\\[1ex]
  \hline
  \hline
  NGC 5128 & 0.0011 & $10^{-22}$ & $0\div10$\\[1ex]
  NGC 1023 Gr.  & 0.0015 & $10^{-22}$ & $0\div10$\\[1ex]
  NGC 2997 & 0.0018 & $10^{-22}$ & $5\div15$\\[1ex]
  NGC 5457 & 0.0019 & $10^{-22}$ & $10\div20$\\[1ex]
  NGC 5033 & 0.0037 & $10^{-22}$ & $25\div35$\\[1ex]
  Virgo  Cl.& 0.0042 & $10^{-22}$ & $30\div40$\\[1ex]
  Fornax Cl. & 0.0044 & $10^{-22}$ & $35\div45$\\[1ex]
  NGC 7582 & 0.0050 & $10^{-22}$& $ 45\div55$\\[1ex]
  Ursa Major Gr. & 0.0057 & $10^{-22}$& $50\div60$\\[1ex]
  Eridanus Cl. & 0.0066 & $10^{-22}$ & $55\div65$\\
  \hline
    \end{tabular}}
\caption {Redshifts, characteristic amplitudes, frequency range for NS-NS systems.}
\label{tab:tabella1}
}
\end{table}

\begin{figure}[]
\centering{
\includegraphics[scale=0.23]{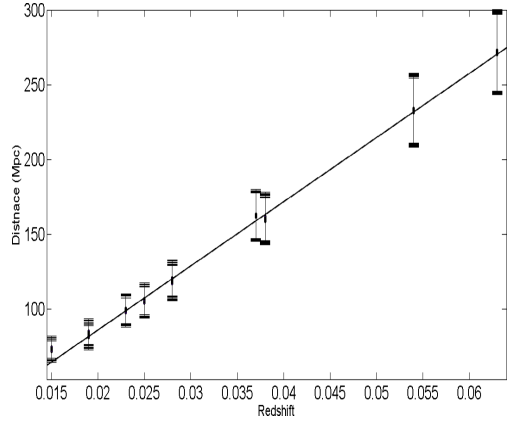}
\caption{Luminosity distance vs redshift for simulated BH-BH systems.}
\label{fig:bh}}
\end{figure}

The Hubble constant value  is $72\pm1$ $km/sMpc$ in agreement with
the recent WMAP  estimation \citep{WMAP7}. The same procedure is adopted for BH-BH systems
($M_C=8.67 M_{\odot}$, $h_C=10^{-21}$). In Tables \ref{tab:tabella2}
 we report the redshift, the value of $h_C$ and the frequency range.  
The simulations are reported in Fig. \ref{fig:bh}, and the Hubble constant value computed by these systems is $69\pm2$
$km/sMpc$.


\begin{table}
\begin{center}
\begin{tabular}{|c|c|c|c|}
  \hline
  \textbf{Object} & \textbf{z} &
 \textbf{$h_c$ } & \textbf{Freq.} \\
& & &  (Hz)\\[1ex]
  \hline
  \hline
  Pavo-Indus  & 0.015& $10^{-21}$& $ 65\div70$\\[1ex]
  Abell 569  & 0.019 & $10^{-21}$ & $ 75\div80$\\[1ex]
  Coma  & 0.023 & $10^{-21}$ & $ 100\div105$\\[1ex]
  Abell 634  & 0.025 & $10^{-21}$ &$ 110\div115$\\[1ex]
  Ophiuchus  & 0.028 & $10^{-21}$ & $ 130\div135$\\[1ex]
  Columba  & 0.034 & $10^{-21}$ & $ 200\div205$\\[1ex]
  Hercules  & 0.037& $10^{-21}$& $205\div210$\\[1ex]
  Sculptor & 0.054 & $10^{-21}$ & $ 340\div345$\\[1ex]
  Pisces-Cetus  & 0.063 & $10^{-21}$ & $ 420\div425$\\[1ex]
  Horologium  & 0.067& $10^{-21}$& $450\div455$ \\[1ex]
  \hline
    \end{tabular}
\caption {For each cluster we indicate redshifts, characteristic amplitudes, frequency range for BH-BH systems.}
\label{tab:tabella2}
\end{center}
\end{table}
%
%
%
%
%
%

\section{Conclusions}

We have considered simulated binary systems
whose redshifts can be estimated considering the barycenter of the
host astrophysical system as  galaxy,  group of galaxies or
cluster of galaxies. In such a way, the standard methods adopted
to evaluate the cosmic distances (e.g. Tully-Fisher or
Faber-Jackson relations) can be considered as "priors" to fit the
Hubble relation.
We have simulated, for example,
NS-NS, and BH-BH binary systems. Clearly, the leading
parameter is the chirp mass $M_c$, or its red-shifted counter-part
${\cal M}_c$, which is directly related to the GW amplitude. The
adopted redshifts are in a  well-tested  range of scales and  the
Hubble constant value is in good agreement with WMAP estimation.
The Hubble-luminosity-distance diagrams of the above simulations
show the possibility to use the coalescing binary systems as
distance indicators and, possibly, as standard candles.  The
limits of the method are, essentially, the measure of GW
polarizations and redshifts. Besides, in order to improve the
approach, a suitable catalogue of observed coalescing
binary-systems is needed. This is the main difficulty of the
method since, being the coalescence a transient phenomenon, it is
very hard to detect and analyze  the luminosity curves of these
systems. Furthermore, a few simulated sources are out of the
LIGO-VIRGO band.

Next generation of interferometer (as LISA\footnote{\cite{LISA}} or
Advanced-VIRGO and LIGO) could play a decisive role to detect GWs
from these systems. At the advanced level, one expects to detect
at least tens NS-NS coalescing events per year, up to distances of
order $2 Gpc$, measuring the chirp mass with a precision better
than $0.1\%$. The masses of NSs are typically of order
$1.4M_\odot$. 
 The most important issue that can be
addressed with a measure of $d_L(z)$ is to understand ``dark
energy'', the quite mysterious component of the energy budget of
the Universe that manifests itself through an acceleration  of the
expansion of the Universe at high redshift. This has been
observed, at $z<1.7$, using Type~Ia supernovae as standard
candles~\citep{Riess,Perl}. A possible concern in these
determinations is the absence of a solid theoretical understanding
of the source. After all, supernovae are complicated phenomena. In
particular, one can be concerned about the possibility of an
evolution of the supernovae brightness with redshift, and of
interstellar extinction in the host galaxy  leading to unknown
systematics. GW standard candles  could lead to completely
independent determinations, and complement and increase the
confidence of other standard candles, \citep{HH}, as well as
extending the result to higher redshifts. In the future, the
problem of the redshift could be obviate finding an
electromagnetic counterpart to the coalescence and short GRBs
could play this role.

In summary, this new type of cosmic distance indicators  could be
considered complementary to the traditional standard candles
opening the doors to a self-consistent {\it gravitational
astronomy}.


\bibliographystyle{aa}

\end{document}